# Hypercomplex Cross-correlation of DNA Sequences

**JIAN-JUN SHU[*] and YAJING LI**

School of Mechanical & Aerospace Engineering,

Nanyang Technological University, 50 Nanyang Avenue, Singapore 639798

## Abstract

A hypercomplex representation of DNA is proposed to facilitate comparison of DNA sequences with fuzzy composition. Using hypercomplex number representation, conventional sequence analysis method, such as, dot matrix analysis, dynamic programming, and cross-correlation method are extended and improved to align the DNA sequences with fuzzy composition. The hypercomplex dot matrix analysis can provide more control over the degree of alignment desired. A new scoring system has been proposed to accommodate the hypercomplex number representation of DNA and integrated with dynamic programming alignment method. By using hypercomplex cross-correlation, matching and mismatching alignment information between two aligned DNA sequences are stored in the resultant real part and imaginary parts, respectively. Mismatching alignment information is particularly useful for refining consensus sequence-based motif scanning.

## 1. Introduction

DNA molecules are composed of four linearly linked nucleotides, adenine (A), thymine (T), guanine (G) and cytosine (C). A DNA sequence can be represented as a permutation of four characters A, T, G, and C at different lengths. The standard symbolic representation of DNA sequences has obvious advantages in the storage, search and retrieval of genomic information, but has certain limitations in handling and manipulating genomic information for pattern matching. Converting DNA sequence into numerical sequence opens the possibility of applying signal processing methods to the analysis of genomic data. There have been several attempts to attach numerical values to the symbolic DNA sequences for different purposes [1-7].

[*] Author to whom correspondence should be addressed.

Cross-correlation is an engineering technique for comparing two numerical signals. When comparing two DNA sequences, two kinds of numerical representation method were used in the literature: One is four-vector encoding by which a DNA sequence is decomposed into four independent binary indicator sequences [1,8]; Another is complex number encoding by which a DNA sequence is converted into one complex number sequence (A, T, G and C are mapped to $1$, $-1$, $i$ and $-i$, respectively) [2]. With using four-vector encoding, four cross-correlations are required to obtain the total matching count between a pair of DNA sequences at each alignment. The disadvantage of this method is that all mismatching alignment information between the two DNA sequences is missed. With complex number encoding, only one cross-correlation is required to obtain matching and mismatching alignment information; however, the real part of the cross-correlation is equal to the number of matches at each alignment of the two DNA sequences minus the number of complementary mismatches (alignment between A and T as well as alignment between G and C). So, the real part of cross-correlation is actually an approximate measure of the similarity between the pair of DNA sequences. The peak in the real part represents the similarity between a pair of DNA sequences, which may disappear due to the offset of complementary mismatches and matches. On the other hand, complex number representation cannot encode DNA base codes with fuzzy composition, such as the IUPAC codes [9] used to represent degenerate consensus sequences. This shortcoming limits the use of cross-correlation to address the problem of DNA consensus sequence matching.

The hypercomplex numbers form a non-commutative four-dimensional number system that extends the commutative one-dimensional real and two-dimensional complex numbers. Since there are four types of nucleotides, a four-dimensional space is essential to represent the DNA codes fully. The hypercomplex number representation of DNA base code proposed in [10] can take the occurrence frequency of each nucleotide in a DNA nucleotide base code into full account. It is very suitable for the numerical encoding DNA base code with fuzzy composition. In this paper, a hypercomplex cross-correlation method in DNA consensus sequence matching analysis is proposed by using the hypercomplex number encoding of DNA sequences.

# 2. Hypercomplex representation of DNA motif

DNA motifs are short, conserved domains that are presumed to have biological functions. Often, they indicate sequence-specific binding sites for proteins such as nucleases and transcription factors. In DNA binding sites, only some positions are conserved among all the sites, and the other positions have a range of variability. So, a DNA binding site is usually represented by a consensus sequence or a position-weight matrix (PWM). As an example, TATA-box consensus sequence, base frequency and PWM for eukaryotic RNA polymerase II promoters are listed in Table 1.

**Table 1** TATA-box consensus sequence, base frequency and PWM for eukaryotic RNA polymerase II promoters [11], and their hypercomplex representation

| **Consensus** | **T** | **A** | **T** | **A** | **W** | **A** | **W** | **R** |
|---|---|---|---|---|---|---|---|---|
| $P_{\mathrm{A}}\,(\%)$ | 4.1 | 90.5 | 0.8 | 91.0 | 68.9 | 92.5 | 57.1 | 39.8 |
| $P_{\mathrm{T}}\,(\%)$ | 79.5 | 9.0 | 96.1 | 7.7 | 31.1 | 1.6 | 31.1 | 8.5 |
| $P_{\mathrm{G}}\,(\%)$ | 4.6 | 0.5 | 0.5 | 1.3 | 0.0 | 5.1 | 11.3 | 40.4 |
| $P_{\mathrm{C}}\,(\%)$ | 11.8 | 0.0 | 2.6 | 0.0 | 0.0 | 0.8 | 0.5 | 11.3 |
| **Position** | **1** | **2** | **3** | **4** | **5** | **6** | **7** | **8** |
| $W_{\mathrm{A}}$ | −3.05 | 0 | −4.61 | 0 | 0 | 0 | 0 | −0.01 |
| $W_{\mathrm{T}}$ | 0 | −2.28 | 0 | −2.34 | −0.52 | −3.65 | −0.37 | −1.4 |
| $W_{\mathrm{G}}$ | −2.74 | −4.28 | −4.61 | −3.77 | −4.73 | −2.65 | −1.5 | 0 |
| $W_{\mathrm{C}}$ | −2.06 | −5.22 | −3.49 | −5.17 | −4.63 | −4.12 | −3.74 | −1.13 |

The consensus sequence 'TATAWAWR' can be represented by the following three different hypercomplex vectors:

1) Consensus sequence-based hypercomplex representation of TATA box:

$$\mathbf{V}_1 = \left(\mathbf{i},\ 1,\ \mathbf{i},\ 1,\ 0.5+0.5\mathbf{i},\ 1,\ 0.5+0.5\mathbf{i},\ 0.5+0.5\mathbf{j}\right),$$

based on the hypercomplex representation of DNA base code [10];

2) Frequency-based hypercomplex representation of TATA box:

$$\mathbf{V}_2 = \begin{pmatrix} 0.041+0.795\mathbf{i}+0.046\mathbf{j}+0.118\mathbf{k},\ 0.905+0.09\mathbf{i}+0.005\mathbf{j},\ 0.008+0.961\mathbf{i}+0.005\mathbf{j}+0.026\mathbf{k}, \\ 0.91+0.077\mathbf{i}+0.013\mathbf{j},\ 0.689+0.311\mathbf{i},\ 0.925+0.016\mathbf{i}+0.051\mathbf{j}+0.008\mathbf{k}, \\ 0.571+0.311\mathbf{i}+0.113\mathbf{j}+0.005\mathbf{k},\ 0.398+0.085\mathbf{i}+0.404\mathbf{j}+0.113\mathbf{k} \end{pmatrix},$$

based on every column of the base frequency, $P_{\mathrm{A}} + P_{\mathrm{T}}\mathbf{i} + P_{\mathrm{G}}\mathbf{j} + P_{\mathrm{C}}\mathbf{k}$, in Table 1;

3) PWM-based hypercomplex representation of TATA box:

$$\mathbf{V}_3 = \begin{pmatrix} -3.05-2.74\mathbf{j}-2.06\mathbf{k},\ -2.28\mathbf{i}-4.28\mathbf{j}-5.22\mathbf{k},\ -4.61-4.61\mathbf{j}-3.49\mathbf{k},\ -2.34\mathbf{i}-3.77\mathbf{j}-5.17\mathbf{k}, \\ -0.52\mathbf{i}-4.73\mathbf{j}-4.63\mathbf{k},\ -3.65\mathbf{i}-2.65\mathbf{j}-4.12\mathbf{k},\ -0.37\mathbf{i}-1.5\mathbf{j}-3.74\mathbf{k},\ -0.01-1.4\mathbf{i}-1.13\mathbf{k} \end{pmatrix}$$

based on every column of the hypercomplex position-weight vector (HPWV), $W_{\mathrm{A}} + W_{\mathrm{T}}\mathbf{i} + W_{\mathrm{G}}\mathbf{j} + W_{\mathrm{C}}\mathbf{k}$, in Table 1.

# 3. Hypercomplex cross-correlation

The cross-correlation is a measure of similarity between two waveforms as a function of a time-lag applied to one of them. It is commonly used to search a long duration signal for a shorter. The cross-correlation is similar in nature to the convolution of two functions. The mathematical form of cross-correlation function can be generalized by using a hypercomplex presentation. Here consider $\mathbf{u} = \left(u_0, u_1, \cdots, u_{N-1}\right)$ and $\mathbf{v} = \left(v_0, v_1, \cdots, v_{N-1}\right)$, two hypercomplex vectors of length $N$. The hypercomplex cyclic cross-correlation operation of $\mathbf{u}$ and $\mathbf{v}$ is defined as the hypercomplex vector $\mathbf{w} = \left(w_0, w_1, \cdots, w_{N-1}\right)$:

$$w_l = \left(\mathbf{u} * \overline{\mathbf{v}}\right)_l = \sum_{n=0}^{N-1} u_{(n+l) \bmod N} \; \overline{v}_n \quad l = 0, 1, \cdots, N-1 \tag{1}$$

where $\overline{v}_n$ is the conjugate of $v_n$. The shift operation on $u_n$ is conducted cyclically using modulo arithmetic for the subtraction.

## 3.1. Hypercomplex pairwise DNA sequence alignment

Hypercomplex cross-correlation is an extension of cross-correlation concept to hypercomplex vector. So, the process of using hypercomplex cross-correlation to compare DNA sequences is similar to the one used in [2]. The difference is that hypercomplex number is used to encode DNA base code: $1$ for A, $\mathbf{i}$ for T, $\mathbf{j}$ for G and $\mathbf{k}$ for C. The algorithm for the pairwise DNA sequence alignment with hypercomplex cross-correlation is as follows, and the schematic diagram is shown in Figure 1.

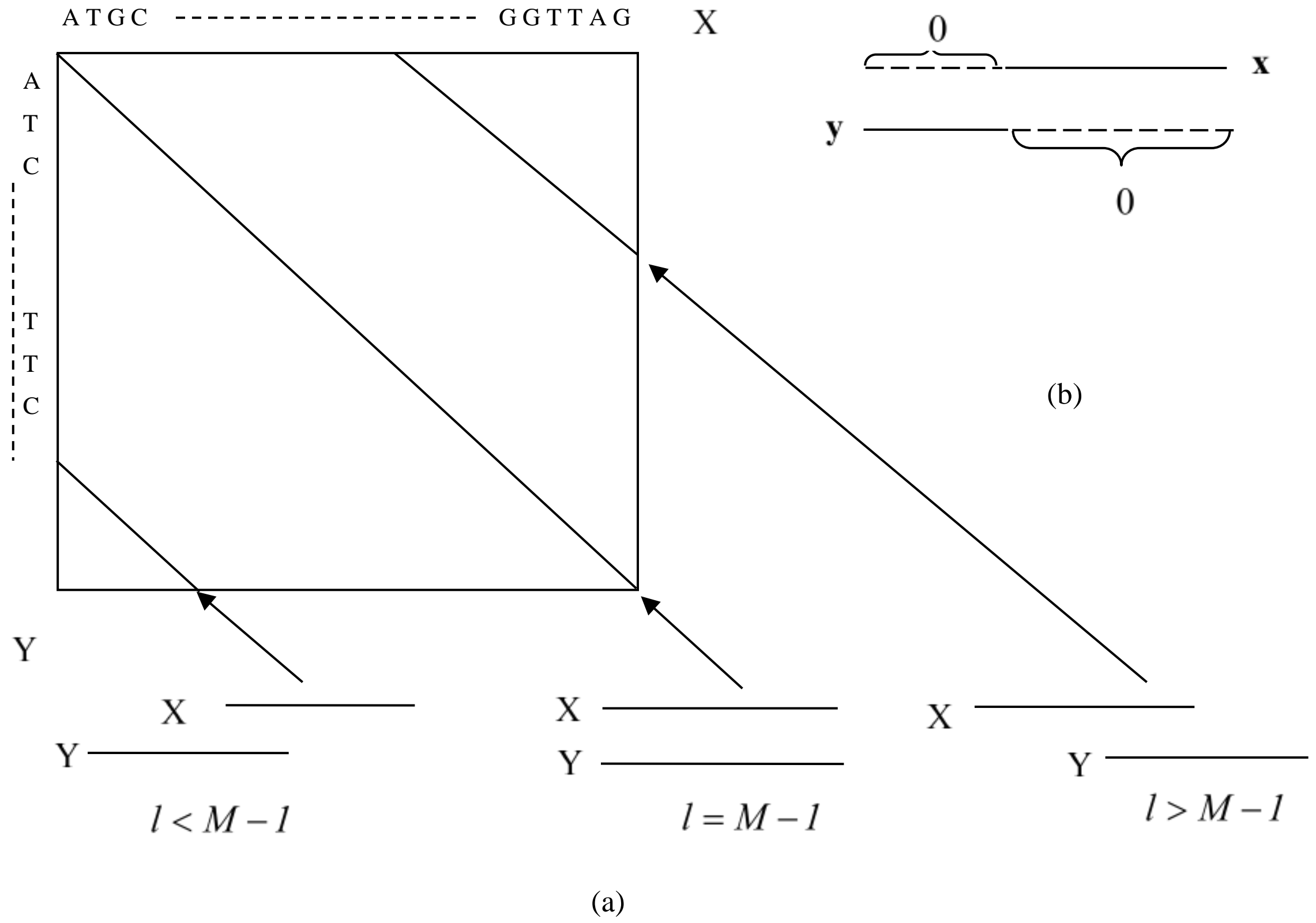

**Figure 1** Schematic diagram of computing sequence similarity by hypercomplex cross-correlation: (a): $\mathrm{X}$ and $\mathrm{Y}$ are two DNA sequences; $l-M+1$ is relative position between $\mathrm{X}$ and $\mathrm{Y}$, position is set when $x_0$ are aligned with $y_0$ as $M-1$ for $l$; real part of hypercomplex cross-correlation is actually equal to matching count of each diagonal; (b): After converting to a hypercomplex vector, pad $\mathbf{x}$ on left side with $M-1$ $0$'s and $\mathbf{y}$ on right side with $N-1$ $0$'s to form two vectors $\mathbf{x} = \left( \overbrace{0,\cdots,0}^{M-1}, x_0, x_1, \cdots, x_{N-1} \right)$ and $\mathbf{y} = \left( y_0, y_1, \cdots, y_{M-1}, \underbrace{0,\cdots,0}_{N-1} \right)$ of length $N+M-1$.

**Algorithm 1**: Given two DNA sequences $\mathrm{X}$ and $\mathrm{Y}$ of length $N$ and $M$, respectively, calculate the pairwise alignment between $\mathrm{X}$ and $\mathrm{Y}$ with hypercomplex cross-correlation:

*Step* 1. Transform the two DNA sequences $\mathrm{X}$ and $\mathrm{Y}$ into hypercomplex vectors $(x_0, x_1, \cdots, x_{N-1})$ and $(y_0, y_1, \cdots, y_{M-1})$.

*Step* 2. Pad $\left(x_0, x_1, \cdots, x_{N-1}\right)$ on the left with $M-1$ $0$'s and $\left(y_0, y_1, \cdots, y_{M-1}\right)$ on the right with $N-1$ $0$'s to form two vectors $\mathbf{x} = \left(\overbrace{0,\cdots,0}^{M-1}, x_0, x_1, \cdots, x_{N-1}\right)$ and $\mathbf{y} = \left(y_0, y_1, \cdots, y_{M-1}, \underbrace{0,\cdots,0}_{N-1}\right)$ of length $N+M-1$.

*Step* 3. Calculate the hypercomplex cyclic cross-correlation $\mathbf{z} = \mathbf{x} * \overline{\mathbf{y}}$ by Equation (1).

*Step* 4. Plot the real part $\mathbf{r} = \left(r_0, r_1, \cdots, r_{N+M-1}\right)$ of $\mathbf{z}$ against the shift $l$. A peak in the real part of hypercomplex cross-correlation indicates similarity between two DNA sequences. The relative position $l-M+1$ corresponding to the maximum of $\mathbf{r}$ is the position that corresponds to the maximal similarity.

When a cross-correlation is performed between two hypercomplex-represented DNA sequences, according to hypercomplex algebra, the hypercomplex product in Equation (1) results in $1$ if the two aligned bases are matched, and $\pm\mathbf{i}$ or $\pm\mathbf{j}$ or $\pm\mathbf{k}$ if they are mismatched (for the detailed results, see Table 2), here $\pm\mathbf{i}$ corresponding to complementary mismatches (A *vs.* T or G *vs.* C); $\pm\mathbf{j}$ to transitional mismatches (A *vs.* G or T *vs.* C); and $\pm\mathbf{k}$ to transversal mismatches (A *vs.* C or G *vs.* T).

**Table 2** Hypercomplex product results between hypercomplex-represented DNA base codes

| | A | T | G | C |
|---|---|---|---|---|
| A | $1$ | $-\mathbf{i}$ | $-\mathbf{j}$ | $-\mathbf{k}$ |
| T | $\mathbf{i}$ | $1$ | $-\mathbf{k}$ | $\mathbf{j}$ |
| G | $\mathbf{j}$ | $\mathbf{k}$ | $1$ | $-\mathbf{i}$ |
| C | $\mathbf{k}$ | $-\mathbf{j}$ | $\mathbf{i}$ | $1$ |

In summary, we have:

$$\begin{cases} d = n_{\mathrm{AA}} + n_{\mathrm{TT}} + n_{\mathrm{GG}} + n_{\mathrm{CC}} \\ a = n_{\mathrm{TA}} + n_{\mathrm{CG}} - n_{\mathrm{AT}} - n_{\mathrm{GC}} \\ b = n_{\mathrm{GA}} + n_{\mathrm{TC}} - n_{\mathrm{AG}} - n_{\mathrm{CT}} \\ c = n_{\mathrm{CA}} + n_{\mathrm{GT}} - n_{\mathrm{AC}} - n_{\mathrm{TG}} \end{cases} \tag{2}$$

where $d$, $a$, $b$ and $c$ denote the real and three imaginary parts of hypercomplex cross-correlation $\mathbf{z}$, $n_{\mathrm{XY}}$ denotes the number of aligned base pair $\mathrm{XY}\left(\mathrm{X}, \mathrm{Y} \in \mathrm{A}, \mathrm{T}, \mathrm{G}, \mathrm{C}\right)$. Thus, the real part of the hypercomplex cross-correlation gives the number of matches per alignment of the two DNA sequences, and the imaginary part reflects the mismatching alignment information: $\mathbf{i}$ imaginary part $a$ is incremented by 1 for each complementary mismatched TA or CG, because of the non-commutativity of the hypercomplex product, $a$ is subtracted by 1 for each mismatched AT or GC; similarly, $\mathbf{j}$ imaginary part $b$ is incremented by 1 for each transitional mismatched GA or TC, $b$ is subtracted by 1 for each mismatched AG or CT; $\mathbf{k}$ imaginary part $c$ is incremented by 1 for each transversal mismatched CA or GT, $c$ is subtracted by 1 for each mismatched AC or TG. Thus, a peak in the real part of the hypercomplex cross-correlation indicates similarity between two DNA sequences. Furthermore, since the real part of hypercomplex cross-correlation only reflects the matching alignment information, the peak value can be used as a measure of the similarity between two DNA sequences: the larger the peak value, the more similar the two DNA sequences are.

## 3.2. Hypercomplex consensus sequence-based motif scanning

When the vector $\mathbf{y}$ is the hypercomplex representation of consensus sequence, the hypercomplex cross-correlation can be used to do consensus sequence-based motif scanning. If the aligned two DNA bases have hypercomplex representations $\mathbf{x} = P_{\mathrm{A}}^{\mathrm{x}} + P_{\mathrm{T}}^{\mathrm{x}}\mathbf{i} + P_{\mathrm{G}}^{\mathrm{x}}\mathbf{j} + P_{\mathrm{C}}^{\mathrm{x}}\mathbf{k}$ and $\mathbf{y} = P_{\mathrm{A}}^{\mathrm{y}} + P_{\mathrm{T}}^{\mathrm{y}}\mathbf{i} + P_{\mathrm{G}}^{\mathrm{y}}\mathbf{j} + P_{\mathrm{C}}^{\mathrm{y}}\mathbf{k}$, where $P$ is the base frequency, the multiplication in Equation (1) is denoted as $R_{\mathrm{xy}}$ and calculated by:

$$
\begin{aligned}
R_{xy} = \mathbf{x}\overline{\mathbf{y}} = {} & P_{\mathrm{A}}^{\mathrm{x}} P_{\mathrm{A}}^{\mathrm{y}} + P_{\mathrm{T}}^{\mathrm{x}} P_{\mathrm{T}}^{\mathrm{y}} + P_{\mathrm{G}}^{\mathrm{x}} P_{\mathrm{G}}^{\mathrm{y}} + P_{\mathrm{C}}^{\mathrm{x}} P_{\mathrm{C}}^{\mathrm{y}} \\
& + ( P_{\mathrm{T}}^{\mathrm{x}} P_{\mathrm{A}}^{\mathrm{y}} + P_{\mathrm{C}}^{\mathrm{x}} P_{\mathrm{G}}^{\mathrm{y}} - P_{\mathrm{A}}^{\mathrm{x}} P_{\mathrm{T}}^{\mathrm{y}} - P_{\mathrm{G}}^{\mathrm{x}} P_{\mathrm{C}}^{\mathrm{y}} )\mathbf{i} \\
& + ( P_{\mathrm{G}}^{\mathrm{x}} P_{\mathrm{A}}^{\mathrm{y}} + P_{\mathrm{T}}^{\mathrm{x}} P_{\mathrm{C}}^{\mathrm{y}} - P_{\mathrm{A}}^{\mathrm{x}} P_{\mathrm{G}}^{\mathrm{y}} - P_{\mathrm{C}}^{\mathrm{x}} P_{\mathrm{T}}^{\mathrm{y}} )\mathbf{j} \\
& + ( P_{\mathrm{C}}^{\mathrm{x}} P_{\mathrm{A}}^{\mathrm{y}} + P_{\mathrm{G}}^{\mathrm{x}} P_{\mathrm{T}}^{\mathrm{y}} - P_{\mathrm{A}}^{\mathrm{x}} P_{\mathrm{C}}^{\mathrm{y}} - P_{\mathrm{T}}^{\mathrm{x}} P_{\mathrm{G}}^{\mathrm{y}} )\mathbf{k}
\end{aligned}
\tag{3}
$$

The real part of $R_{\mathrm{xy}}$ is the probability of finding a match in the alignment, since the hypercomplex number representation assigned to each DNA nucleotide base code is based on the base frequency $P$ at which each base occurs in the base code. The imaginary part of $R_{\mathrm{xy}}$ reflects the probability of finding a mismatch in the alignment: the $\mathbf{i}$ imaginary part is the probability of finding TA or CG pair minus the probability of finding AT or GC pair; the $\mathbf{j}$ imaginary part is the probability of finding GA or TC pair minus the probability of finding AG or CT pair; the $\mathbf{k}$ imaginary part is the probability of finding CA or GT pair minus the probability of finding AC or TG pair. Thus, the real part of the hypercomplex cross-correlation $\mathbf{z}$ gives a score that can be used to measure the similarity between DNA substring and the consensus sequence. The algorithm for the DNA motif scanning with hypercomplex cross-correlation is as follows:

**Algorithm 2**: Given a consensus sequence $\mathrm{Y}$ of length $M$ and a DNA sequence $\mathrm{X}$ of length $N$, scan DNA sequence $\mathrm{X}$ to identify all the similar segment of $\mathrm{X}$ to the consensus sequence $\mathrm{Y}$.

*Step* 1. Transform the two DNA sequences $\mathrm{X}$ and $\mathrm{Y}$ into hypercomplex vectors $\left(x_0, x_1, \cdots, x_{N-1}\right)$ and $\left(y_0, y_1, \cdots, y_{M-1}\right)$.

*Step* 2. Pad $(x_0, x_1, \cdots, x_{N-1})$ on the left with $M-1$ $0$'s and $(y_0, y_1, \cdots, y_{M-1})$ on the right with $N-1$ $0$'s to form two vectors $\mathbf{x} = \left( \overbrace{0,\cdots,0}^{M-1}, x_0, x_1, \cdots, x_{N-1} \right)$ and $\mathbf{y} = \left( y_0, y_1, \cdots, y_{M-1}, \underbrace{0,\cdots,0}_{N-1} \right)$ of length $N+M-1$.

*Step* 3. Calculate the hypercomplex cyclic cross-correlation $\mathbf{z} = \mathbf{x} * \overline{\mathbf{y}}$ by Equation (1).

*Step* 4. Plot the real part $\mathbf{r} = (r_M, r_{M+1}, \cdots, r_N)$ of $\mathbf{z}$ against $l$. Position $l - M + 1$ corresponding to the maximum value of $\mathbf{r}$ is the starting position when the DNA substring matches the consensus sequence (potential binding site).

Hypercomplex cross-correlation can be implemented by the direct evaluation of the summation formula in Equation (1), which requires running time $O(N^2)$. The cost of performing direct evaluation of cross-correlation can be reduced to $O(N \ln N)$ by the hypercomplex Fourier transform.

## 3.3. Hypercomplex PWM-based motif scanning

If the vector $\mathbf{y}$ is changed to the hypercomplex representation of PWM, the hypercomplex cross-correlation can be used for profile matching [12,13]. The real part of multiplication in Equation (1) is $W_{\mathrm{A}}^{\mathrm{x}} W_{\mathrm{A}}^{\mathrm{y}} + W_{\mathrm{T}}^{\mathrm{x}} W_{\mathrm{T}}^{\mathrm{y}} + W_{\mathrm{G}}^{\mathrm{x}} W_{\mathrm{G}}^{\mathrm{y}} + W_{\mathrm{C}}^{\mathrm{x}} W_{\mathrm{C}}^{\mathrm{y}}$. Thus, the real part of the hypercomplex cross-correlation $\mathbf{z}$ is equal to the profile matching score.

## 3.4. Hypercomplex Fourier transform (HFT)

A hypercomplex number can be represented in the polar form by generalizing the Euler's formula for complex numbers. That is:

$$\mathbf{z} = d + a\mathbf{i} + b\mathbf{j} + c\mathbf{k} = |\mathbf{z}| \mathrm{e}^{\boldsymbol{\mu}\theta} = |\mathbf{z}| (\cos\theta + \boldsymbol{\mu} \sin\theta) \tag{4}$$

where

$$\cos\theta = \frac{d}{|\mathbf{z}|}, \quad \sin\theta = \frac{\sqrt{a^2 + b^2 + c^2}}{|\mathbf{z}|}, \quad |\mathbf{z}| = \sqrt{d^2 + a^2 + b^2 + c^2} \tag{5}$$

then $\boldsymbol{\mu}$ can be expressed as

$$\boldsymbol{\mu} = \frac{a}{\sqrt{a^2 + b^2 + c^2}} \mathbf{i} + \frac{b}{\sqrt{a^2 + b^2 + c^2}} \mathbf{j} + \frac{c}{\sqrt{a^2 + b^2 + c^2}} \mathbf{k} \tag{6}$$

$$\boldsymbol{\mu}^2 = -1$$

$\boldsymbol{\mu}$ is a unit pure hypercomplex number, and it is referred to the eigen-axis. Note that $\boldsymbol{\mu}$ represents the direction in the three-dimensional space of imaginary part of hypercomplex number and $\theta$ is referred to the eigen-angle.

The discrete HFT is generalized from discrete complex Fourier transform (FT). Due to the non-commutativity of hypercomplex number multiplication, HFT is defined as:

$$Z(\kappa) = \mathrm{HFT}\{z_n\} = \frac{1}{\sqrt{N}} \sum_{n=0}^{N-1} z_n \, \mathrm{e}^{-\boldsymbol{\mu}_0 \frac{2\pi}{N} \kappa n} \tag{7}$$

Its inverse discrete hypercomplex Fourier transform (IHFT) is given by

$$z_n = \mathrm{IHFT}\{Z(\kappa)\} = \frac{1}{\sqrt{N}} \sum_{\kappa=0}^{N-1} Z(\kappa) \, \mathrm{e}^{\boldsymbol{\mu}_0 \frac{2\pi}{N} \kappa n} \tag{8}$$

where the vector $\boldsymbol{\mu}_0$ is called the axis of the transform, and it is an arbitrary unit pure hypercomplex number. For computing convenience, the vector $\mathbf{i}$ is chosen as the transform axis here after.

## 3.5. Implementation of hypercomplex Fourier transform (HFT) by Fourier transform (FT) and inverse Fourier transform (IFT)

Assume the input hypercomplex series $z_n = d_n + a_n\mathbf{i} + b_n\mathbf{j} + c_n\mathbf{k}$ and a given transform axis is $\mathbf{i}$, HFT $Z(\kappa)$ is calculated as:

$$Z(\kappa) = \mathrm{HFT}\{z_n\} = \frac{1}{\sqrt{N}} \sum_{n=0}^{N-1} z_n \, \mathrm{e}^{-\mathbf{i} \frac{2\pi}{N} \kappa n} = \mathrm{FT}\{z_n^{(1)}\} + \mathrm{IFT}\{z_n^{(2)}\} \, \mathbf{j} \tag{9}$$

where $z_n = z_n^{(1)} + z_n^{(2)} \mathbf{j}$, $z_n^{(1)} = d_n + a_n\mathbf{i}$ and $z_n^{(2)} = b_n + c_n\mathbf{i}$. Assume two hypercomplex vectors $\mathbf{x}$ and $\mathbf{y}$, hypercomplex cross-correlation can be calculated by the following equation [14]:

$$z_n = \mathrm{IHFT}\{Z(\kappa)\} = \frac{1}{\sqrt{N}} \sum_{\kappa=0}^{N-1} Z(\kappa) \, \mathrm{e}^{\mathbf{i} \frac{2\pi}{N} \kappa n} = \mathrm{IHFT}\left\{\mathrm{HFT}\{x_n\} \overline{\mathrm{HFT}\{y_n^{(1)}\}} - \mathrm{IHFT}\{x_n\} \mathrm{IHFT}\{y_n^{(2)}\} \, \mathbf{j}\right\}. \tag{10}$$

By using HFT, the computational complexity of the Algorithm 2 is reduced to $O(N \ln N)$ [14,15].

# 4. Results and discussion

## 4.1. Hypercomplex pairwise alignment of DNA

Three human immunodeficiency virus (HIV) sequences are used to demonstrate the application of hypercomplex cross-correlation for DNA pairwise alignments. There are two human type 1 isolates HIVMN (HIV type 1, isolate MN, GenBank accession number M17449) and HIVRF (HIV type 1, isolate RF, GenBank accession number M17451), and one simian virus, SIVMM142 (Simian immunodeficiency virus isolate MM142m-83, GenBank accession number Y00277). Each sequence is between 9000 and 10,000 bases in length and available in GenBank [16]. Figure 2(a) shows the real part of the hypercomplex cross-correlation of HIVMN with HIVRF. The horizontal axis is the relative shift of HIVRF with respect to HIVMN. The vertical axis is the number of matches. Figure 2(b) is similar to Figure (a) but shows the real part of the hypercomplex cross-correlation of HIVMN with SIVMM142.

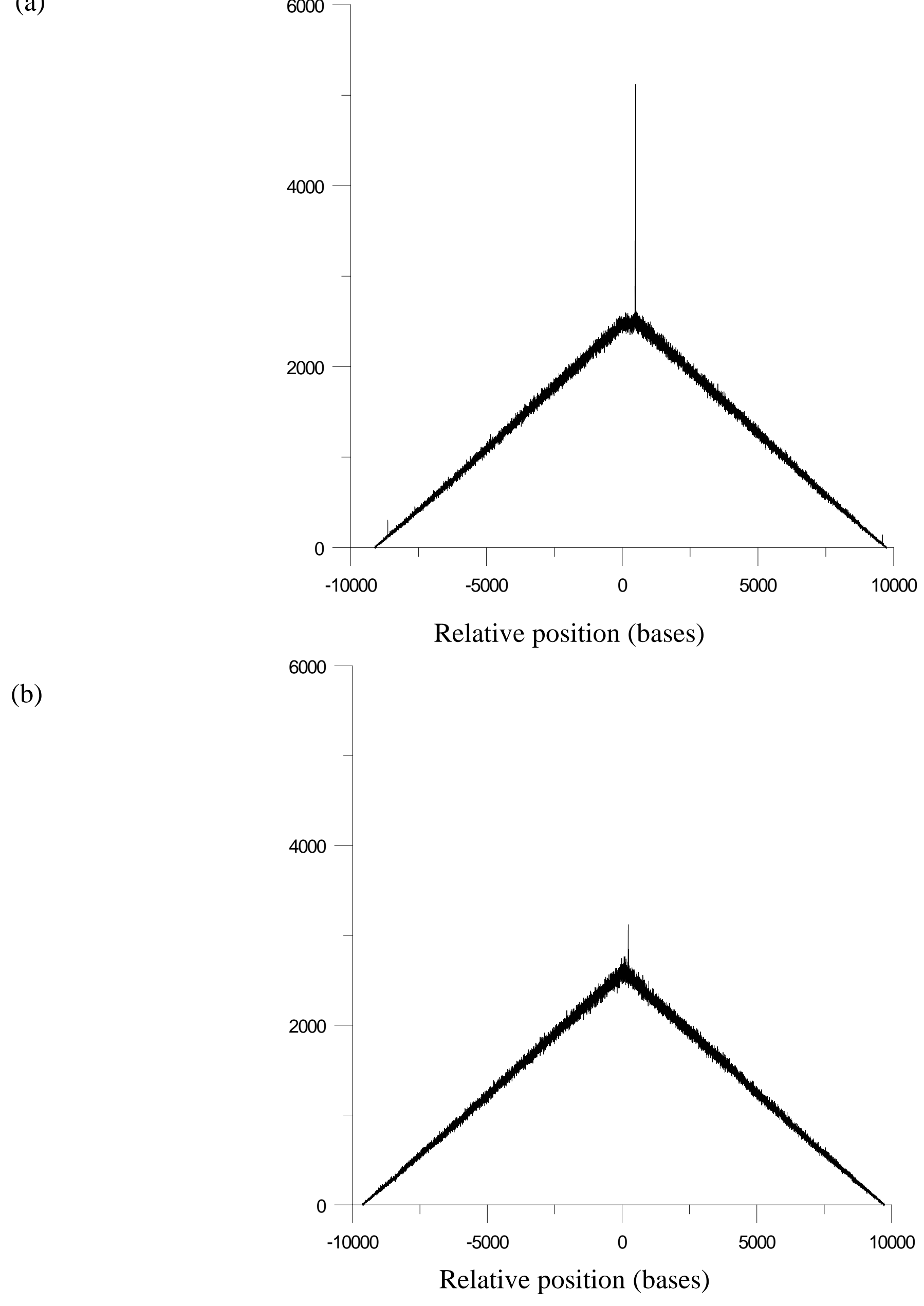

**Figure 2** Real part of hypercomplex correlation of (a) HIVMN *vs.* HIVRF and (b) HIVMN *vs.* SIVMM142

A peak in the real part of the hypercomplex cross-correlation indicates similarity between two compared sequences. Peak value can be used to measure similarity, for example, as expected, the peak near zero shift, is much larger in the comparison of two different HIV isolates than in the comparison of SIV and HIV isolates.

## 4.2. Hypercomplex consensus sequence-based motif scanning

Another application of hypercomplex cross-correlation is to scan a query DNA sequence for the potential transcription binding sites. An example of using hypercomplex correlation to locate potential TATA boxes in a query sequence for the H. sapiens H4/g gene (gi number: 32003) is given here. TATA-box consensus sequence, base frequency and PWM for eukaryotic RNA polymerase II promoters and their corresponding hypercomplex representation are listed in Table 1. From experiment, the TATA signal locates at position 185 as substring 'TATTTAAG'. When scanning for potential TATA boxes in the H. sapiens H4/g gene using conventional string-matching method [17], no exact match to the consensus sequence 'TATAWAWR' is found; however, when using the cross-correlation with hypercomplex encoding, potential TATA boxes can be found by setting a reasonable score threshold.

The real part of hypercomplex cross-correlation of H. sapiens H4/g gene against the consensus sequence 'TATAWAWR' is shown in Figure 3. If the score threshold is set as 5.5, two potential TATA boxes are identified to be located at position 185 and 381 of the query DNA sequence, respectively. Then the best candidate should be identified to show most similar to consensus sequence. One simply way is to line the candidate substring against the consensus sequence and then perform the base-by-base comparison. This operation can be expressed in an alignment graph. For the candidate at position 185 (candidate 1: 'TATTTAAG'), the alignment graph reads as follows:

```
T A T T T A A G
| | |   | | | |
T A T A W A W R
```

For the candidate at position 381 (candidate 2: 'TATCTATG'), the alignment graph is:

```
T A T C T A T G
| | |   | | | |
T A T A W A W R
```

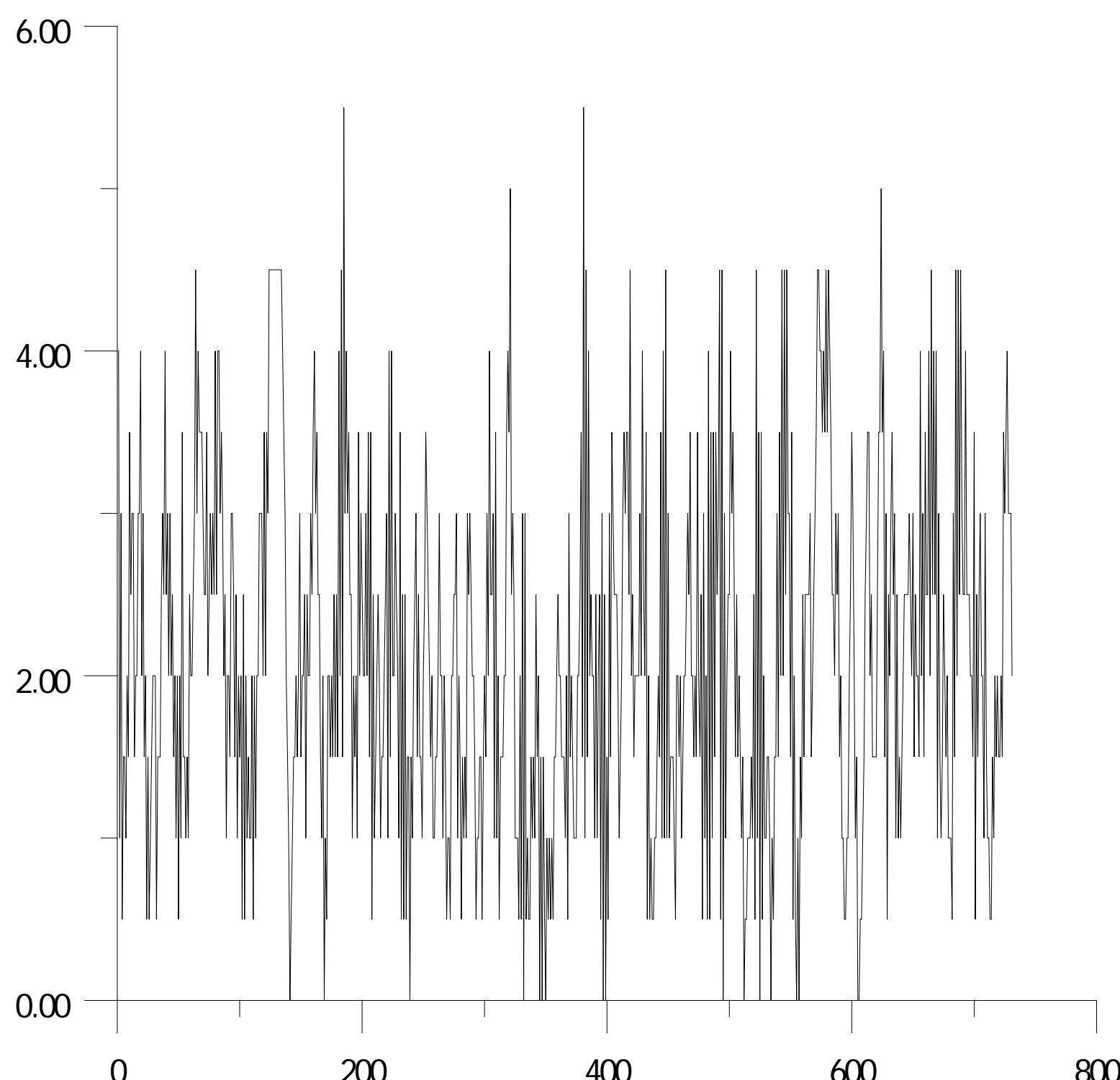

**Figure 3** Real part of hypercomplex cross-correlation of H4/g gene against consensus sequence ‘TATAWAWR’. A hypercomplex representation of consensus sequence is a vector $\mathbf{V}_1$ in Table 1.

From the alignment plot, both candidates match the consensus sequence at all positions except position 4. For candidate 1, the mismatch is a complementary mutation of ‘A to T’; for candidate 2, the mismatch is a transversal mutation ‘A to C’. Since complementary mismatches occur more probably than transversal mismatches in the binding site, candidate 1 is more similar to the consensus sequence. Thus, the substring at position 185 is the best candidate.

A second approach to identify the best candidate is to use mismatching alignment information from hypercomplex cross-correlation. For each peak in the Figure 2, the correlation shift value $l$ is determined, hold $l$ fixed and plot in Figure 4 the partial sum function $S_l(m)$:

$$S_l(m) = \sum_{n=0}^{m-1} x_{(n+l) \bmod N} \overline{y_n} \quad m = 1, 2, \cdots, N \tag{11}$$

The real part of the partial sum reflects matching alignment information, and the imaginary parts reflect mismatching alignment information. For the real part graph (line 1), a region with a slope of +1 indicates a perfect match, a horizontal region corresponds to a mismatch, and a region with a slope less than 1 corresponds to a fuzzy match (for example alignment between T and W). For the $\mathbf{i}$ imaginary part plot (line 2), a region with a slope of +1 or -1 indicates a perfect complementary mismatch, a horizontal region indicates no complementary mismatch in that region, and a region with a slope less than 1 corresponds to a fuzzy mismatch. The similar meaning is for **j** and **k** imaginary part plots (lines 3 and 4).

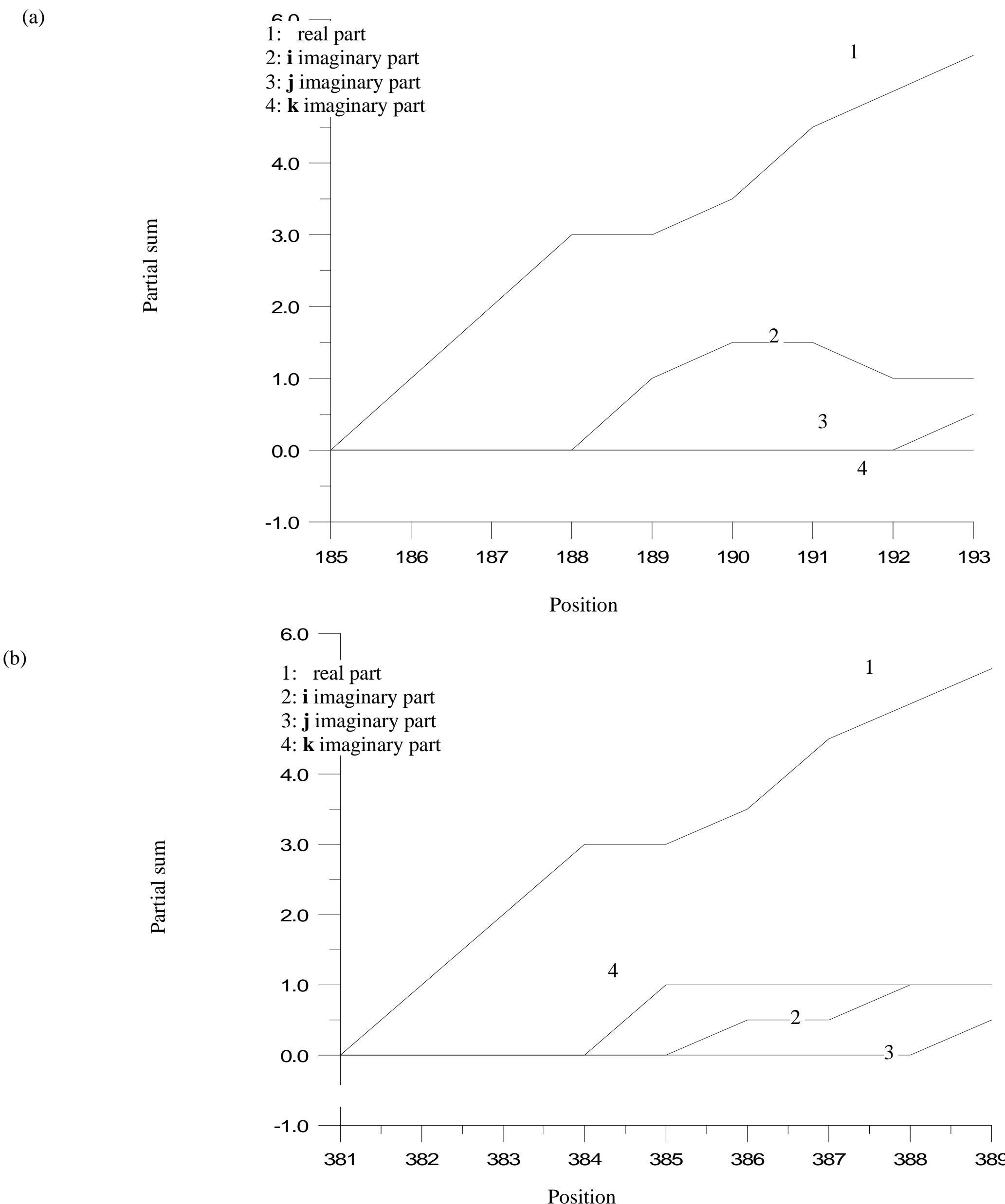

**Figure 4** Partial sum plots of H. sapiens H4/g gene compared to consensus sequence 'TATAWAWR', showing matched and mismatched alignment information for positions (a) 185 and (b) 381.

Figure 4 shows that the mismatch is located at position 4 of each candidate substring: for candidate 1, the mismatch is a complementary mutation; for candidate 2, the mismatch is a transversal mutation. Thus, the candidate at position 185 is the best candidate.

Another approach to identify the best candidate is to use the hypercomplex encoding of TATA boxes based on the base frequency at each position (the vector $\mathbf{V}_2$ shown in Table 1). The real part of hypercomplex cross-correlation is shown in Figure 5. The candidate at position 185 can be easily identified as the best candidate since it corresponds to the maximal score of the real part.

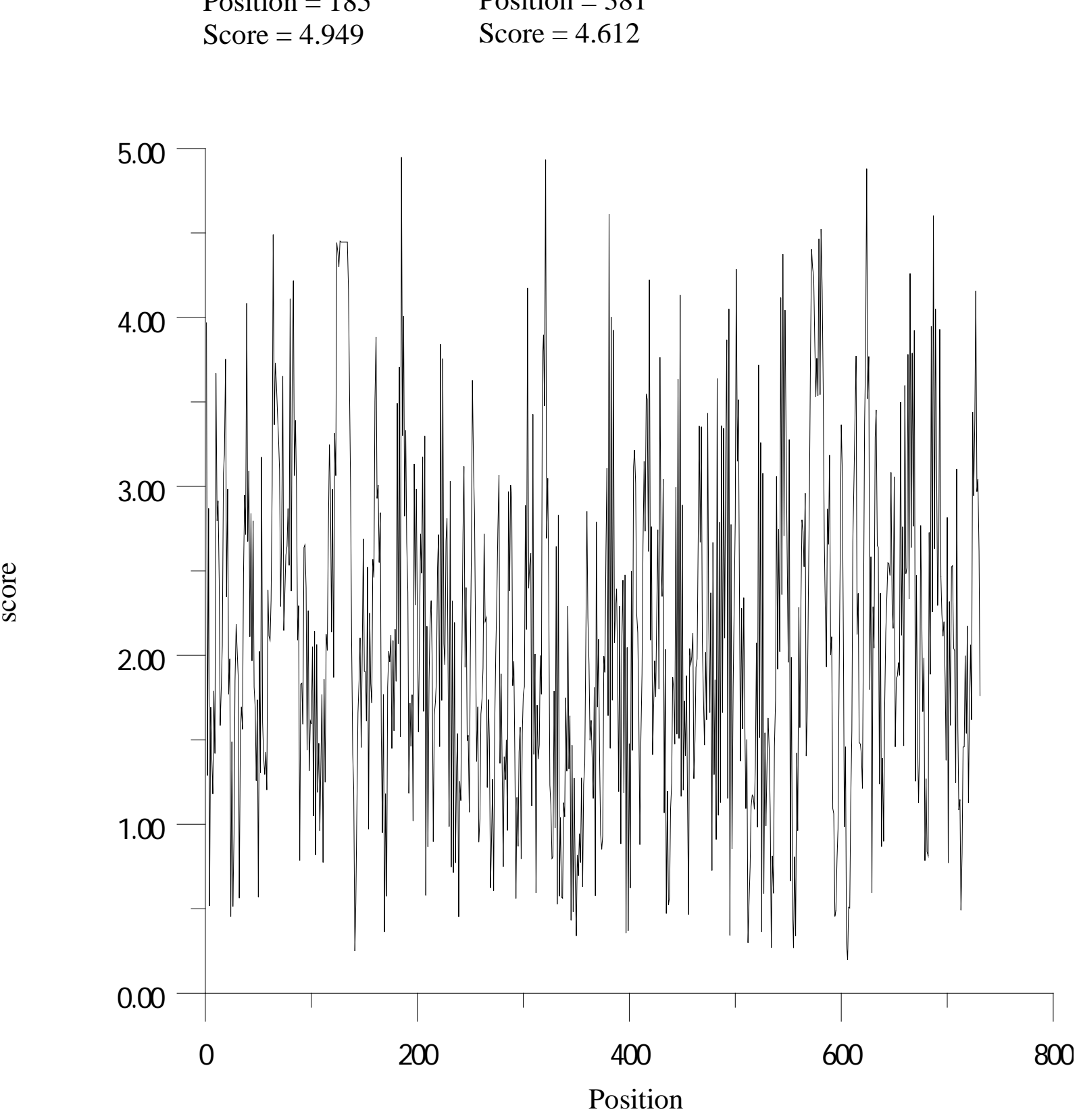

**Figure 5** Real part of hypercomplex cross-correlation of H. sapiens H4/g gene against frequency-based consensus sequence. A hypercomplex representation is a vector $\mathbf{V}_2$ in Table 1.

## 4.3. Hypercomplex PWM-based motif scanning

Finally, the real part of hypercomplex correlation result of H. sapiens H4/g gene against the HPWV (the vector $\mathbf{V}_3$ shown in Table 1) is shown in Figure 6. The real part of the hypercomplex correlation result is equal to the profile matching score. The substring can easily be identified at position 185 as the most probably TATA box.

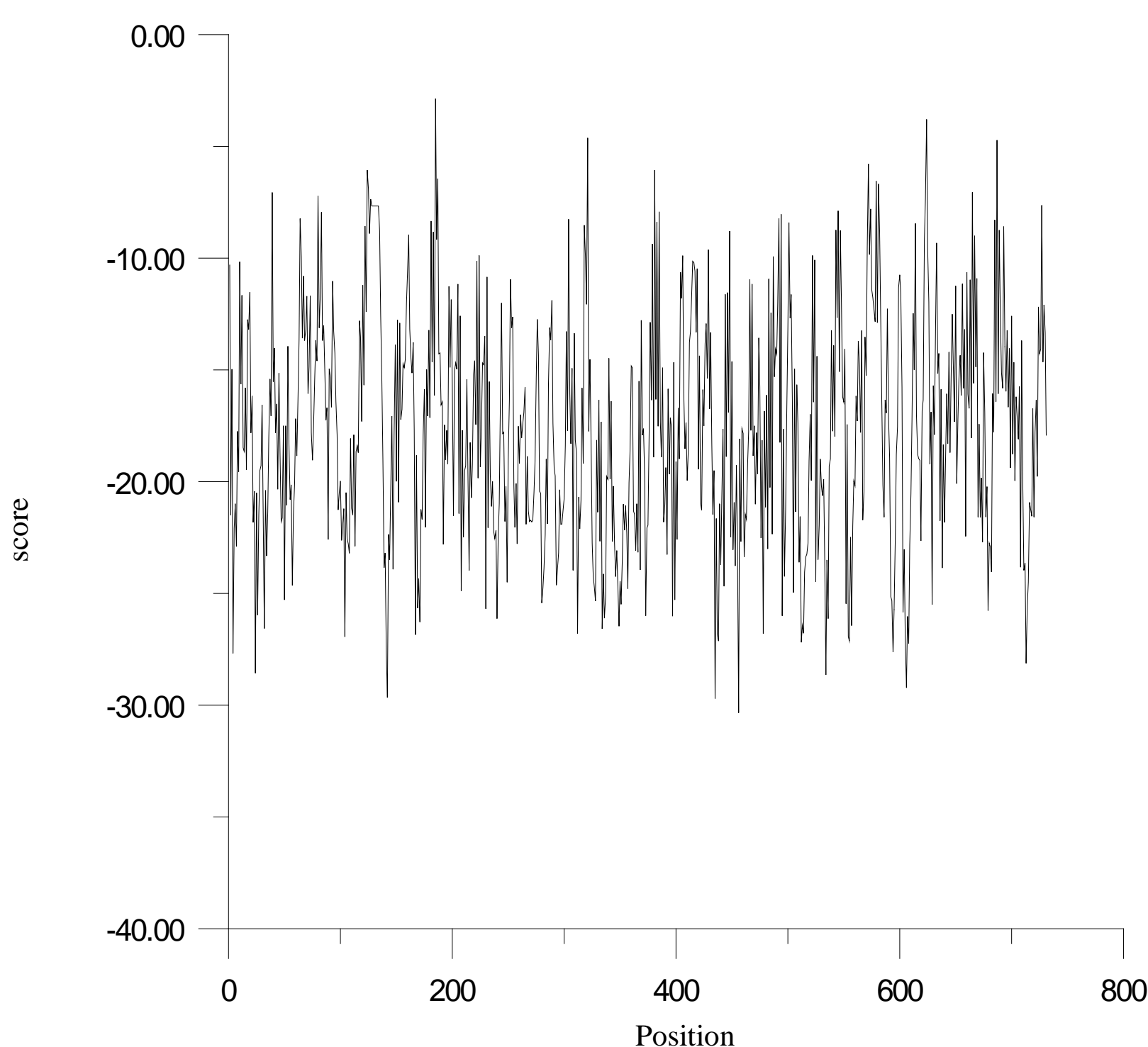

**Figure 6** Real part of hypercomplex cross-correlation between H. sapiens H4/g gene and hypercomplex position weight vector $\mathbf{V}_3$ is shown in Table 1.

# 5. Conclusion

The conventional way to handle motif scanning is to search for good matching substring alignments, such as, the real-number-based fast Fourier transform approach [8] and the hypercomplex-number-based dynamic programming matrix approach [10]; however, these approaches, solely based on matching alignment information, are susceptible to a large number of ambiguous or mispredicted sites, especially if the motif sequence is poorly conserved. To reduce the number of false positives, the mismatched components of the alignment need to be considered. Using the hypercomplex representation of DNA, matching and mismatching alignment information between two compared DNA sequences are separated and stored in the real part and imaginary parts of the cross-correlation result, respectively. So, the real part of cross-correlation is an exact measure of the similarity between the pair of DNA sequences at each alignment. Moreover, by emphasizing the probabilistic occurrence of each DNA base at a specific position through the hypercomplex representation of DNA, cross-correlation method can be used to scan the query DNA sequence for potential transcription binding sites, the real part of the hypercomplex cross-correlation gives a score indicating the similarity degree of the DNA substring to the consensus sequence. With a reasonable score threshold, this approach is able to find binding site candidates that cannot be detected by conventional string-matching method. When combined with partial sum graphs, hypercomplex cross-correlation is capable of not only detecting the presence of potential binding sites, but also locating the position of the best candidate from the mismatching alignment information of the compared DNA sequences, enabling consensus sequence matching as accurate as profiling matching [18-22]. Furthermore, the hypercomplex cross-correlation is computationally efficient when implemented with the FFT, with computational complexity scaling as $O(N \ln N)$, making this method potentially useful for large computationally intensive tasks, such as database searching.